# Bacteriorhodopsin: Tunable Optical Nonlinear Magnetic Response

F.A.Bovino [1], M.C.Larciprete [2], C.Sibilia [2], M.Giardina [1], G. Váró [3], C.Gergely [4,5]

1. Quantum Optics Lab-Selex-SI, Via Puccini – Genova, Italy Tel. +39-010-658-2841, fbovino@selex-si.com.
2. Dipartimento SBAI- Universita' di Roma La Sapienza – Via Scarpa 16-00161 Roma,Italy
3. Institute of Biophysics, Biological Research Center of the Hungarian Academy of Sciences, H-6701 Szeged, Hungary
4. Université Montpellier 2, Laboratoire Charles Coulomb UMR 5221, F-34095, Montpellier,France
5. CNRS, Laboratoire Charles Coulomb UMR 5221, F-34095, Montpellier, France

**Abstract:** We report on a strong and tunable magnetic optical nonlinear response of Bacteriorhodopsin (BR) under "off resonance" femtosecond (fs) pulse excitation, by detecting the polarization map of the noncollinear second harmonic signal of an oriented BR film, as a function of the input beam power. BR is a light-driven proton pump with a unique photochemistry initiated by the all trans retinal chromophore embedded in the protein. An elegant application of this photonic molecular machine has been recently found in the new area of optogenetics, where genetic expression of BR in brain cells conferred a light responsivity to the cells enabling thus specific stimulation of neurons. The observed strong tunable magnetic nonlinear response of BR might trigger promising applications in the emerging area of pairing optogenetics and functional magnetic resonance imaging susceptible to provide an unprecedented complete functional mapping of neural circuits.

---

**1. Introduction**

Natural optical activity of bacteriorhodopsin (BR), a robust and abundant chromoprotein, has been recently elegantly valorized in optogenetics: internalization of opsin genes into neurons confers light responsiveness to these cells enabling specific light stimulation of neural circuits. This application motivated us to further investigate the intimate optical properties of BR behaving as a photonic molecular machine. Here we show a strong and tunable magnetic optical nonlinear response of BR under "off resonance" femtosecond pulse excitation, by detecting the polarization map of the noncollinear second harmonic signal as a function of the input beam power. The observed strong nonlinear magnetization of BR might trigger promising applications in the emerging area of pairing optogenetics and functional magnetic resonance imaging susceptible to provide an unprecedented complete functional mapping of neural circuits.

In the realm of fluorescent probes, application of bioluminescent proteins like the green fluorescence protein, has contributed to the recent outstanding advances in biological imaging [1,2]. Nonlinear optical response of certain proteins opened the avenue for cellular and tissue imaging based on the emission of second harmonic (SH) signal of these endogen probes, monitored by two-photon microscopy. Examples are the green fluorescence protein and the bacteriorhodopsin (BR), where the SH signal was related to the induced dipole of the chromophores within the proteins [3].

An elegant application of the BR chromoprotein has been recently found in the new area of optogenetics, where genetic expression of BR in brain cells conferred a light responsivity to the cells enabling thus specific stimulation of neurons [4]. Recently, research has been reported on combination possibilities of optogenetics with functional magnetic resonance imaging (fMRI) for direct monitoring of the optically stimulated brain activities [5]. All these findings initiated us to investigate the intimate optical properties of the bacteriorhodopsin, a robust and naturally abundant protein, including its tunable optical nonlinear magnetic response, yet never been studied to the best of our knowledge.

BR is a light-driven proton pump with a unique photochemistry initiated by the all trans retinal chromophore embedded in the protein. The nonlinear magnetic response of BR is related to its chiral properties. It is well known that chiral molecules can exhibit nonlinear magnetic response at optical frequencies. Molecular chirality is of very high importance in biology, chemistry and materials science [6]. Chirality is equivalent to the absence of mirror rotation axes of any order and it can be evidenced by optical means due to the optical activity of the molecules [7]. The interaction of chiral molecules with light is sensitive to the circular polarization, furthermore the plane of polarization of linearly polarized light can rotate, however optical activity is a linear optical effect. A model to describe the chirality is founded on the electron path in the molecule: as the electrons of chiral molecules are displaced from their equilibrium by the application of the electromagnetic field, they are forced to move along helical like paths. This gives rise to an induced magnetic dipole moment of the molecule in addition to the electric dipole moment, therefore chiral molecules respond to both the electric and magnetic component of the field. Nonlinear optical properties of molecules, such as second harmonic generation (SHG) and sum frequency generation (SFG), have been shown to be very sensitive to chirality [7,8]. The nonlinear response is typically detected by the optical rotatory dispersion ( SHG-ORD), i.e. the rotation angle of the $\hat{s}$ -polarized component with respect to $\hat{p}$ -polarized one or by detecting the circular dichroism [9]. Materials with non linear optical properties (NLO) can be found also among biological molecules presenting an ordered arrangement and non-centrosymmetric structure in space, or having a chromophore. The latter can be found in chromoproteins, best represented by bacteriorhodopsin (BR). The transmembrane protein bacteriorhodopsine (BR) is one of the simplest known active membrane transport system [9,10]. It functions as a ligth driven proton pump converting ligth energy into a proton gradient across the bacterial cell membrane [10,11]. Trimers of BR proteins are placed in a hexagonal two dimension lattice within the purple membrane, forming thereby a real crystalline structure (Figures 1). Furthermore, each BR monomer contains a covalently bound retinal chromophore, presenting its own transition dipole, which is responsible for its outstanding quadratic nonlinear optical response [12] enhanced by the protein environment [13], as well as for the light absorption [14] in the visible range. However, one of the most intriguing properties of BR is the chirality [9]. In what follow we describes an interesting BR behaviour related to ist chiral structure.

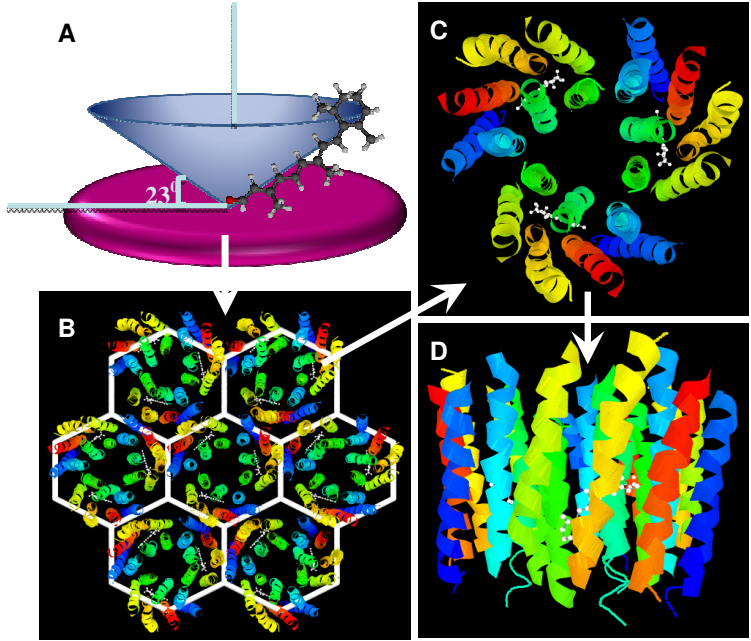

*Figure 1: A) In the BR the cromophore is oriented at ~ 23° with respect to the plane of the purple membrane (from ref. 26). B) Hexagonal arrangement of the BR timers in the purple membrane - P3 symm.class. C) Upper view of the unit cell of BR trimers where the retinal chromophores (in white) are visible. D) One BR is organized in 7 helices and a retinal cromophore buried inside the protein*

## 2. Experimental Results

The nonlinear polarization at 2ω induced by the optical E(ω) electromagnetic field consists of a series of multipole terms [15]: $\vec{P}_{eff}(2\omega) = \vec{P}_D(2\omega) - i\vec{k}\vec{Q}(2\omega) - \frac{1}{\omega}\vec{k} \times \vec{M}(2\omega)$ where $\vec{P}_D, \vec{Q}, \vec{M}$ represent the electric–dipole, electric-quadrupole and magnetic-dipole polarization, respectively. The electric dipole contribution is proportional to the nonlinear electrical susceptibility $\chi^{eee}$ tensor and to the magnetic dipole $\chi^{eem}$ tensor: $\vec{P}_D(2\omega) = \sum_{j,k} \chi^{(2)eee}_{ijk} E_j(\omega)E_k(\omega) + \chi^{(2)eem}_{ijk} E_j(\omega)B_k(\omega)$. The $\chi^{eem}$ tensor has nonzero elements and similar to surface contributions, it is originated by a bulk contribution, i.e. depending on the coherence length, when a two beam excitation is performed [15], unless the medium from which the fundamental beam is incident has a nonvanishing nonlinearity. The magnetic contribution to the second order nonlinear polarization is
$$\vec{M}(2\omega) = \sum_{j,k} \chi^{(2)mee}_{ijk} E_j(\omega)E_k(\omega).$$

Nonlinear magnetization is typically detectable from the surface of chiral molecules [16], and it has the same order of magnitude like the electric dipole contribution for Langmuir-Blodgett films of chiral polythiophene [17]. Finally, the quadruple contribution is typically negligible with respect to the other contributions in a homogeneous bulk material.
Magnetic dipole interaction and/or interference between electric and magnetic dipoles have been considered to contribute to the nonlinear response of chiral BR, together with the intrinsic chirality and chromophore orientation [18], however to discriminate the above mentioned contributions is far to be trivial.

We found an evident role of the different sources of nonlinear polarization and in particular the nonlinear magnetic one, by using a method based on noncollinear second harmonic generation [19,20]. We performed measurements on an oriented BR film prepared by using the asymmetric electrostatic interaction of the surface charge of the membrane fragments with a charged support surface. We employed an electrophoretic deposition technique to grow a 13 μm thick oriented film onto a substrate covered by a 60nm thick ITO film [18]. The resulting BR films, composed by ~2600 purple membrane layers (of 5nm thickness each) were characterized in terms of homogeneity, optical and electrical properties. The chromophore retinal axis is oriented at an angle of $23 \pm 4°$ with respect to the plane of the purple membrane [21] (fig. 1D) so to form an isotropic conical polar distribution around the normal.

To avoid the optical absorption, the experiments have been performed by using an off resonance excitation: the output of a mode-locked femtosecond Ti:Sapphire laser system tuned at ω=830 nm (76 MHz repetition rate, 130 fs pulse width) was split into two beams of about the same intensity. The polarization of both beams was varied with two identical rotating half wave plates in the range -90°-+90° degrees; $\hat{p}$-polarized pump beam is obtained when $\varphi_1$ and $\varphi_2$ are both 0°, and $\hat{s}$-polarized beams when polarization angles of both pumps are set to ± 90°. Thus the resulting experimental curves of generated signal describe all possible combinations for the polarization state of the two pump beams. The sample was placed onto a rotation stage which allowed to vary the sample rotation angle, α, with a resolution of 0.05 degrees. After passing trough two collimating lenses of 150 mm focal length, the pump beams were sent to intersect in the focus region with the angles β=3.5° and γ= -3.5°, measured with respect to α=0°. Thus for a given α≠ 0°, the corresponding incidence angles of the two pump beams result to be $\alpha_1=\alpha-\beta$ and $\alpha_2=\alpha-\gamma$, respectively (fig. 2A).

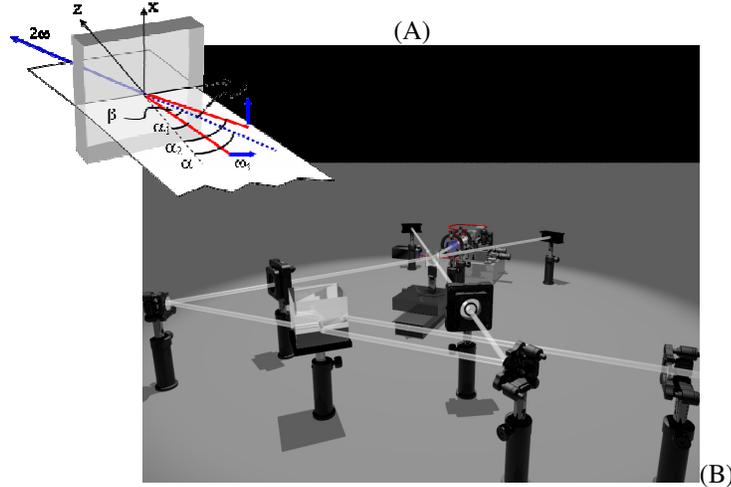

*Figure 2 – (A) The two input beams impinges with a fixed relative angle and the noncollinear SH signal is detected along the bisector axis forming an angle of 40° with respect to the normal incidence; (B) experimental set up showing the directions of the fundamental beams.*

Finally, the temporal overlap of the incident pulses was automatically controlled with an external delay line. The experimental apparatus is shown in fig.2B. The SH signal emitted by the BR sample is approximately along the bisector of the aperture angle between the two pump beams, whatever the sample rotation angle, and an analyzer allows to select the desired SH polarization state.

Figure 3 show the experimental $\hat{s}$-polarized SH signal from BR under an average input power of 50mW (fig.3A) and then changing the average input power of the Fundamental Field (FF) for different values up to 750 mW (fig 3F). We observe a gradual increase of the SH signal (up to two orders of magnitude) and a change of the shape of the chart: the maximum of the $\hat{s}$-polarized SH signal is mostly located in the negative quadrant region of $\phi_1$ an $\phi_2$ at low power, it moves toward the positive quadrant increasing the power, saturating the change of position on the polarization chart at 750 mW. The maximum value of the SHG signal follows the parabolic dependence of the intensity, however the location of the maximum SH signal on the polarization chart as a function of the input beam intensity shows an unexpected strong nonlinear dependence of the polarization state of the two input beams. The polarization chart of the $\hat{p}$-polarized SH is only weakly sensitive to input field changes, therefore in the following we do not report those charts, but only the $\hat{s}$-polarized SH signal.

The observed effect is similar to the intensity dependent polarization rotation appearing in third order nonlinear materials [22, 23], i.e. occurring at the same pumping frequency, but in our experiments BR responds with a second order nonlinear polarization at the 2ω frequency and the form of the tensor's elements are responsible of the observed behaviour.

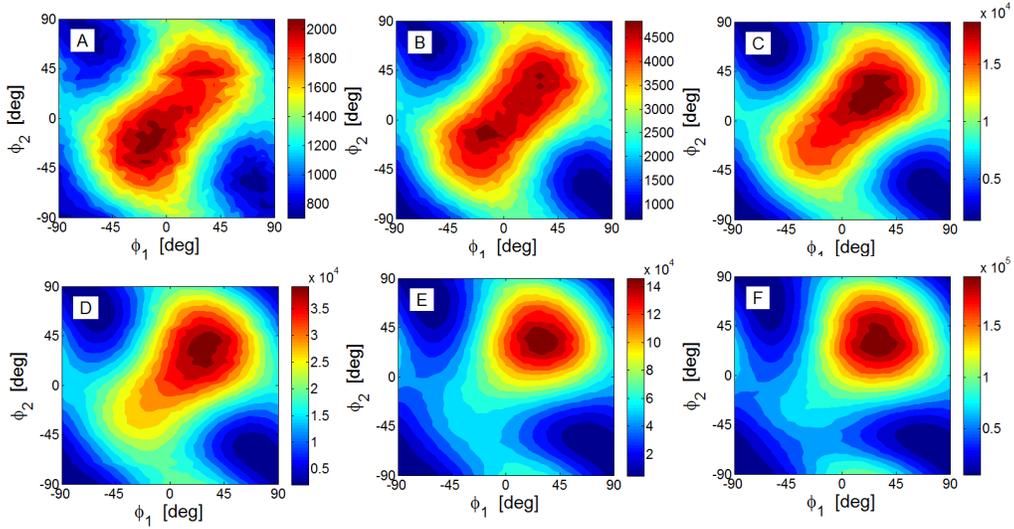

*Figure 3- Experimental $\hat{s}$-polarized SH signal map of BR with a relative angle of FF beam at 7°, with the sample rotated of 40° and for different input powers : A) 50 mW , B) 125 mW, C) 250 mW, D) 375 mW, E) 625 mW and F) 750 mW.*

## 4. Discussion of Results

The full expression of the generated SH power is a function of sample rotation angle $\alpha$, as well as propagation angle and polarization state of both fundamental and generated beams, through the Fresnel coefficients and includes the effective susceptibility, $\chi^{(2)}_{eff}(\alpha)$ [24], which is the projection of the tensor on the suitable SH output polarization state. In our case is given by

$$\chi^{(2)}_{eff} = (\hat{p}_{out})(\chi_{il}^{(eee)})(\hat{p}_{in}^{1\,e})(\hat{p}_{in}^{2\,e}) + \\ (\hat{p}_{out})(\chi_{il}^{(eem1)})(\hat{p}_{in}^{1\,e})(\hat{p}_{in}^{2\,m}) + (\hat{p}_{out}^{e})(\chi_{il}^{(eem2)})(\hat{p}_{in}^{1\,m})(\hat{p}_{in}^{2\,e}) + \\ (\hat{p}_{out})(\chi_{il}^{(mee)})(\hat{p}_{in}^{1\,e})(\hat{p}_{in}^{2\,e})$$

(1)

being $\hat{p}_{out}$ is the polarization versor of the SH signal, $\hat{p}_{in}^{1}$ is the polarization versor of the input pump signal and $\hat{p}_{in}^{2}$ is the one of the second pump beam defined in the laboratory frame. The eee, eem, mee denominators stand for the electric dipole, magnetic dipole components of the second order susceptibility tensor $\chi$. To recover the experimental results, we need to include in the model the electric dipole contribution $\chi_{ijk}^{(2)eee}$, the magnetic dipole elements of the tensor $\chi_{ijl}^{(2)eem}$ and the nonlinear magnetization $\vec{M}(2\omega) = \sum_{j,k} \chi_{ijk}^{(2)mee} E_j(\omega) E_k(\omega)$, where the $\chi_{ijk}^{(2)mee}$ and $\chi_{ijl}^{(2)eem}$ elements follow the same symmetry rules of electric dipole elements. However $\chi_{eff}^{mee}$ is in phase-quadrature with respect to $\chi_{eff}^{eee}$, as it comes from the full expression of the nonlinear polarization. As pointed out in refs.[9,28], the magnetic dipole tensor $\chi_{ijl}^{(2)eem}$ is not zero when a two beams interaction is performed, and it contributes twice to the total $\chi_{eff}^{(2)}$, i.e. we get

$$\chi_{ijk}^{(2)eem} \vec{E}_j(\omega) \vec{B}_k(\omega) = i\chi_{ijk}^{eem1} : (\vec{k}_1 x \vec{E}_1)\vec{E}_2 + i\chi_{ijk}^{eem2} : \vec{E}_1(\vec{k}_2 x \vec{E}_2)$$, being $\vec{k}_1$, $\vec{k}_2$ the wave vector of each FF beam. In what follows $\chi_{eff}^{eem}$ and $\chi_{eff}^{mee}$ are the projections of the corresponding tensor giving the $\hat{s}$-polarized signals.

BR has the point group symmetry P3, under Kleinman's symmetry rule far from optical resonances its second order susceptibility tensor, $\chi_{ijk}$, has three nonvanishing components [26,27], i.e. along with the piezoelectric contraction, $\chi_{15}=\chi_{24}$, $\chi_{31}=\chi_{32}$ and $\chi_{33}$. Two additional nonzero components of the nonlinear susceptibility tensor, $\chi_{14} = -\chi_{25}$, determine the so-called chiral contribution to the nonlinear optical response; in fact they appear only if molecules have no planes of symmetry.

Each of the above mentioned nonvanishing elements are present in all the nonlinear tensors responsible of the electric –electric dipole, electric-magnetic dipole and magnetic contribution to the second harmonic signal. The elements of the electric dipole tensor fulfil the above mentioned relations, while the magnetic dipole tensor contains the additional term $\chi_{36}$ and there is a little difference between the $\chi_{31}$ and $\chi_{32}$ magnetic dipole elements. With our measurements we do not evaluate the absolute value of the nonlinear coefficients, but we've fixed the relations among them.

Finally, for the $\hat{s}$-polarized SH signal we have the following expression of the nonlinear coefficients

$$\chi_{eff}^{\phi_1\phi_2\to s} = \left(\chi_{eff}^{(eee)_s}\right) + \left(\chi_{eff}^{(eem1)_s}\right) + \left(\chi_{eff}^{(eem2)_s}\right) + \left(\chi_{eff}^{(mee)_s}\right) \quad (2)$$

The nonvanishing elements of each tensor $\chi_{ijk}$ are related to the hyerpolarizability of the molecule, i.e. to $\beta_{i'j'k'}^{eee}$, $\beta_{i'j'k'}^{eem}$, $\beta_{i'j'k'}^{mee}$, that is the nonlinear response of molecules at the microscopic level in their coordinate system. The macroscopic response is found by averaging over the microscopic response for the various orientations of the molecule in space with respect to the laboratory coordinate system [29].

For a chiral molecule $\beta_{i'j'k'}$ can be expressed as function of the radius $\rho$ and the pitch $\zeta$ [29], if a phenomenological model based on the electron moving along an helical path is adopted. To retrieve the experimental SH signal we have then introduced a relative change of $\rho$ and $\zeta$, according to $\zeta = \zeta_\circ \pm \Delta\zeta = \zeta_\circ(1 \pm \frac{\Delta\zeta}{\zeta_0}), \rho = \rho_\circ \pm \Delta\rho = \rho_\circ(1 \pm \frac{\Delta\rho}{\rho_0})$, where the variation has been fixed proportional to the input power.

By considering each nonlinear contribution to the SH signal as a separate source, and plotting the polarization map for each of them, we obtain the electric dipole $\chi_{eff}^{eee}$ map (fig. 4A): it follows the expected behaviour typical of a symmetry class P3, but this term alone cannot retrieve the experimental one. The magnetic-dipole contributions $\chi_{eff}^{eem1}$ and $\chi_{eff}^{eem2}$ are shown in figures 4B and 4C. The magnetic nonlinear tensor $\chi_{eff}^{mee}$ has a maximum value located in the part of the chart where polarization angles have positive values (fig 4D).

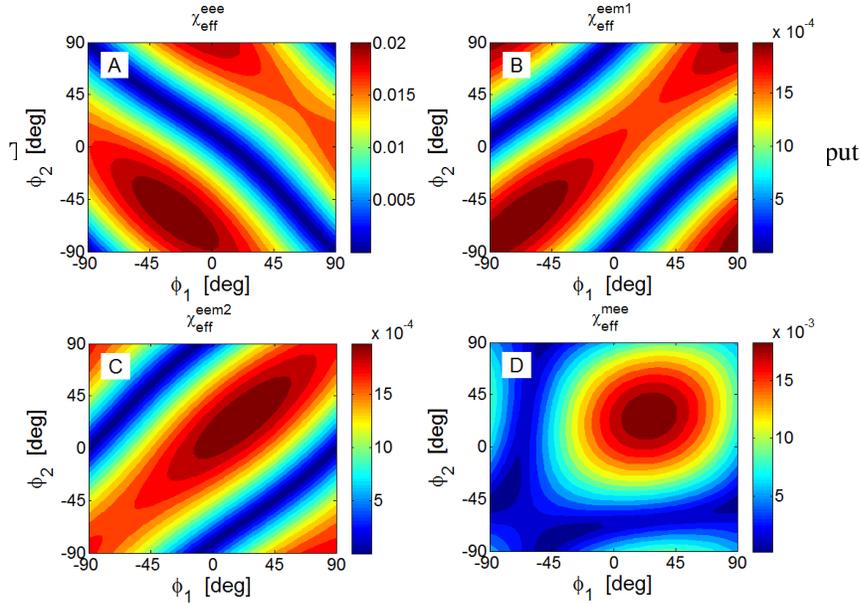

*Figure 4- Calculated polarization chart of nonlinear tensors responsible of $\hat{S}$ -polarized SHG generated by BR : A) $\chi_{eff}^{eee}$, B) $\chi_{eff}^{eem1}$, C) $\chi_{eff}^{eem2}$, D) $\chi_{eff}^{mee}$.*

Figures 5 reveal the total map of the effective susceptibility $\chi^{(2)}_{eff}$ responsible of $\hat{s}$-polarized SH field changing the input power. The trend of SH signal as a function of the input power is in agreement with what we calculate from the phenomenological model based on the single electron model as discussed in ref.[29]. Electrons are assumed to move along a helicoidal path, a change of dimensions of the path proportional to the input power, although the excitation is at off resonance, modifies the value of elements of the tensors, so that the relative weight of the electric dipole and magnetic contributions is varied. In this way the tensors map shape does not change, only their values increase. At higher power, magnetic elements $\chi^{mee}_{eff}$ seem to dominate the $\hat{s}$-polarized SH: the main contributions come from $\chi^{mee}_{31}$ and $\chi^{mee}_{33}$ elements, while for the electric dipole and magnetic dipole sources $\chi^{eee}_{14}$ and $\chi^{eee}_{15}$ dominate the nonlinear response. This becomes evident from fig. 5F, where the $\chi^{(2)}_{eff}$ map at higher power is shown. Any change of the above mentioned tensor elements induces a complete change of the chart. Passing from low power to high power, magnetic dipole contributions appear to be competitive with the electric dipole contribution and finally at higher power it is evident that the source of the nonlinearity is the magnetic dipole. The generated signal from the above mentioned tensor map follows the experimental values in terms of location of maxima and of amplitude.

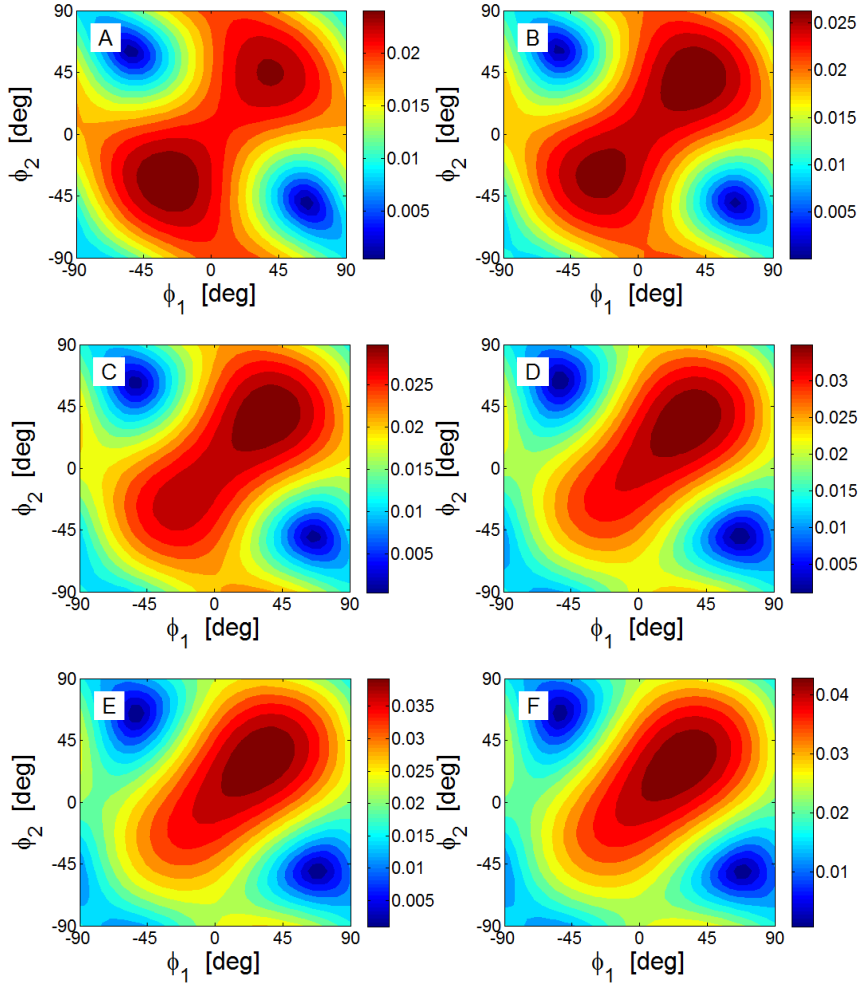

*Figure 5 -Calculated effective susceptibility for different input powers A) 50 mW , B) 125 mW, C) 250 mW, D) 375 mW, E) 625 mW , F) 750 mW.*

## 5. Conclusions

The obtained results reveal a very important behaviour of BR: a change of input power, although off resonance affects the magnetic second order nonlinear response rendering it detectable for $\hat{s}$-polarized SH, moreover the observed magnetic response is of the same order of magnitude as the electric dipole contribution. The utilized method is very efficient to put into evidence the nonlinear response of BR, the different contributions to the SH signal and it reveals a tunability of the nonlinear magnetic response as a function of the input power.

In conclusion, we have evidenced a strong nonlinear magnetic response of BR, by means of noncollinear SHG monitoring the $\hat{s}$-polarized signal. For the best of our knowledge these results constitute the first observation of a tunable nonlinear magnetic response of BR. This finding might revolutionize and further enlarge the application fields of this protein especially in domains where magnetic response is important. Bacteriorhodopsin is already successfully used for optogenetics. The reported magnetic response of BR at light activation makes it an ideal candidate for use in integrating optogenetics with fMRI for global mapping of neural activities and pathways within specific brain regions. If fMRI lit up in the same places where the BR marked neurons were stimulated, one could be confident that fMRI was really picking up brain activation.

## Acknowledgments

C.S. and MC.L.would like to thank Prof. M. Bertolotti for helpful discussions and comments. The work has been partially supported by MARINE project, Italian Ministry of Defence.